\documentclass[epj]{svjour}
\usepackage[usenames]{color}
\usepackage{float}
\usepackage{graphicx}
\usepackage{tikz}
\usetikzlibrary{calc,fadings,decorations.pathreplacing,backgrounds,plotmarks,shapes.geometric}
\usepackage{amssymb}
\usepackage{stackrel}
\usepackage{amsmath}
\usepackage[english]{babel}
\usepackage{float}
\usepackage{rotating}

\begin{document}

\title{Dissipation and energy balance in electronic dynamics of Na clusters}
\titlerunning{Dissipation and energy balance in cluster dynamics}
\author{
M. Vincendon\inst{1,2}
\and
E. Suraud\inst{1,2}
\and
P.-G. Reinhard\inst{3}\fnmsep\thanks{\email{paul-gerhard.reinhard@fau.de}}
}
\institute{
Institut f\"ur Theoretische Physik, Universit\"at
Erlangen, D-91058 Erlangen, Germany
\and
Universit\'e de Toulouse; UPS; Laboratoire de Physique Th\'eorique
  (IRSAMC); F-31062 Toulouse, France
\and
CNRS; LPT (IRSAMC); F-31062 Toulouse, France
}

\date{First draft: 31. 12. 2016}

\abstract{ We investigate the impact of dissipation on the energy
  balance in the electron dynamics of metal clusters excited by strong
  electro-magnetic pulses. The dynamics is described theoretically by
  Time-Dependent Density-Functional Theory (TDDFT) at the level of
  Local Density Approximation (LDA) augmented by a self interaction
  correction term and a quantum collision term in Relaxation-Time
  Approximation (RTA). We evaluate the separate contributions to the
  total excitation energy, namely energy exported by electron
  emission, potential energy due to changing charge state, intrinsic
  kinetic and potential energy, and collective flow energy.  The
  balance of these energies is studied as function of the laser
  parameters (frequency, intensity, pulse length) and as function of
  system size and charge. We also look at collisions with a highly
  charged ion and here at the dependence on the impact parameter
  (close versus distant collisions). Dissipation turns out to be small
  where direct electron emission prevails namely for laser frequencies
  above any ionization threshold and for slow electron extraction in
  distant collisions. Dissipation is large for fast collisions and at
  low laser frequencies, particularly at resonances.  }


\maketitle

\section{Introduction}

Time-Dependent Density-Functional Theory (TDDFT) is the starting point
and the leading tool to simulate the dynamics of many-fermion systems,
in electronic systems \cite{Gro90,Gro96,Mar12} as well as in nuclei
\cite{Neg82aR,Dav85a,Ben03aR}. The Local Density Approximation (LDA)
provides a robust and efficient mean-field description of dynamics
which allows to cover a huge range of phenomena from the linear regime
of small-ampli\-tude oscillations (also known as random-phase
approximation) \cite{Ber94aB} to systems possibly highly excited by
strong laser pulses \cite{Rei06aR,Fen10} or hefty collisions
\cite{golabek2009,Obe10a}. However, more detailed observations and/or
long-time evolution is often sensitive to all sorts of many-body
correlations beyond the mean-field approach \cite{Rei94aR}. A
particularly important class are dynamical correlations from
two-fermion collisions. They add dissipation to the mean-field motion
which has important consequences in a great variety of dynamical
scenarios and systems, e.g., for collisional broadening of excitation
spectra \cite{Ber83aR}, for necessary thermalization steps in nuclear
reactions \cite{Abe96,Sar12a,Lac14a}, for thermalization in highly
excited electronic systems \cite{Del98a,Voi00}.  Dissipation (and
thermalization) is a particularly important and much discussed process
in the dynamics of small metal clusters, see e.g.
\cite{Nae97,Cam00,Voi00,Sch01c,Feh06b,Kje10}.  In the present paper,
we address dissipation and energy transport in small metal clusters
taking up an affordable approach to dissipation, the Relaxation-Time
Approximation (RTA), which had been implemented recently for
simulations of finite electronic systems \cite{Rei15d}.

Although highly desirable, theoretical investigations of dissipation
in finite fermion systems have been hampered so far by the enormous
computational demands for a microscopic description of two-body
collisions in the quantum regime.  The way from the full many-body
hierarchy down to a mean-field description augmented by dynamical
correlations has been thoroughly developed since long in classical
non-equilibrium thermodynamics \cite{Rei98aB}, leading eventually to
the much celebrated Boltzmann equation to account for dynamical
correlations in classical systems \cite{Cer88}. A manageable scheme
for a fully quantum mechanical description in finite systems is still
a matter of actual research. One important quantum feature is the
Pauli principle. It can be accounted for by extending the Boltzmann
collision term to the Boltzmann-Uehling-Uhlenbeck (BUU) form
\cite{Ueh33}.  This semi-classical BUU approach (also known as
Vlasov-Uehling-Uhlenbeck (VUU) equation) provides an acceptable
picture at sufficiently large excitations where quantum shell effects
can be ignored. It has been extensively used in nuclear physics
\cite{Ber88,Dur00} and also employed for the description of metal
clusters in a high excitation domain \cite{Dom98b,Fen04}.  Although
very successful, BUU/VUU is valid only for sufficiently high
excitation energies. And even in the high-excitation domain,
de-excitation by ionization can quickly evacuate large amounts of
excitation energy thus cooling the system down into a regime where
quantum effects count again dominantly. In any case, there is an
urgent need for a quantum description augmented by relaxation effects.

Such dissipative quantum approaches are still well manageable in
bulk systems and have been extensively studied in the framework of
Fermi liquid theory \cite{Kad62}.  It was found that global features
of dissipation can often be characterized by one dominant, exponential
relaxation mode.  This motivated the Relaxation Time Approximation
(RTA) as was introduced in \cite{Bha54} and later on applied to a wide
variety of homogeneous systems \cite{Ash76,Pin66}.  The quantum case
for finite systems is much more involved. A full description of
detailed correlations has been carried through in schematic model
systems \cite{Dut12a} and in the time-dependent
configuration-interaction (TD-CI) method \cite{Kra07a}, both being
nevertheless limited to simple systems. A stochastic treatment of the
quantum collision term promises a tractable approach \cite{Rei92b}.
It has meanwhile been successfully tested in one-dimensional model
systems \cite{Sur14d,Lac16a} and will be developed further.  Recently,
RTA has been been successfully implemented as dissipative extension of
TDLDA for finite systems and applied to the realistic test case of Na
clusters \cite{Rei15d}. This now provides an affordable and efficient
approach to dissipation in finite fermion systems.

The present paper uses RTA to study systematically the dynamics of Na
cluster during and after laser excitation in dependence on the key
laser parameters, frequency, intensity, and pulse length.  At the side
of observables, we concentrate here on the energy balance.  To this
end we introduce the various contributions to the excitation energy,
namely intrinsic kinetic and potential energy, charging energy, and
energy loss by electron emission. 
The paper is organized as follows. 
In section \ref{sec:frame}, we summarize the numerical handling of
TDLDA and the RTA scheme.
In section, \ref{sec:energies} we introduce in detail the key
observables used in this study, the various contributions to the energy.
In section \ref{sec:results}, we present the results, especially the
energy balance as function of the various laser parameters.
Further technical details are provided in appendices.

\section{Formal framework}
\label{sec:frame}

\subsection{Implementation of TDDFT}

Basis of the description is mean-field dynamics with Time-Dependent
Density Functional Theory (TDDFT). Actually, we employ it at the level
of the Time-Dependent Local-Density Approximation (TDLDA) treated in
the real time domain \cite{Gro90,Gro96}.  It is augmented by a
Self-Interaction Correction (SIC) approximated by average-density SIC
(ADSIC) \cite{Leg02} in order have correct ionization potentials
\cite{Klu13}, which is crucial to simulate electron emission properly.
The time-dependent Kohn-Sham equations for mean field and
single-electron wave functions are solved with standard techniques
\cite{Cal00,Rei04aB}.  The numerical implementation of TDLDA is done
in standard manner~\cite{Cal00,Rei04aB}.  The coupling to the ions is
mediated by soft local pseudopotentials~\cite{Kue99}.  The electronic
exchange-correlation energy functional is taken from Perdew and
Wang~\cite{Per92}. 

The Kohn-Sham potential is handled in the
Cylindrically Averaged Pseudo-potential Scheme (CAPS)
\cite{Mon94a,Mon95a}, which has proven to be an efficient and reliable
approximation for metal clusters close to axial symmetry.
Wavefunctions and fields are thus represented on a 2D cylindrical grid
in coordinate space \cite{Dav81a}.  For the typical example of the
Na$_{40}$ cluster, the numerical box extends up to 104 a$_0$ in radial
direction and 208 a$_0$ along the $z$-axis, while the grid spacing is
0.8 a$_0$. To solve the (time-dependent) Kohn-Sham equations for the
single particle (s.p.) wavefunctions, we use time-splitting for time
propagation~\cite{Fei82} and accelerated gradient iterations for the
stationary solution \cite{Blu92}. The Coulomb field is computed with
successive over-relaxation \cite{Dav81a}.  We use absorbing boundary
conditions~\cite{Cal00,Rei06c}, which gently absorb all outgoing
electron flow reaching the boundaries of the grid. The difference between
the initial number of electrons and the actual number of electrons
left in the simulation box is thus  a measure for ionization in terms
of $N_\mathrm{esc}$, the number of escaped electrons.

The external laser field is described as a classical electro-magnetic wave in the
long wavelengths limit. This augments the Kohn-Sham Hamiltonian by
a time-dependent external dipole field
\begin{eqnarray}
\label{eq:Elaser}
  U_\mathrm{ext}(\mathbf{r},t) 
  &=& 
  e^2\mathbf{r}\cdot\mathbf{e}_\mathrm{z} 
  \, E_0 \, \sin(\omega_\mathrm{las}t) f(t)
  \quad,
\\
  f(t)
  &=&
  \sin^2\left(\pi\frac{t}{T_\mathrm{pulse}}\right)
  \theta(t)\theta(T_\mathrm{pulse}-t)
  \quad.
\end{eqnarray}
The laser features therein are: the (linear) polarization
$\mathbf{e}_\mathrm{z}$ along the symmetry axis, the peak field
strength $E_0$ related to laser intensity as $I_0\propto E_0^2$, the
photon frequency $\omega_\mathrm{las}$, and the total pulse length
$T_\mathrm{pulse}$. The latter is related to the full width at half
maximum (of intensity) as FWHM = $T_\mathrm{pulse}/3$.

This basic building block, mean-field propagation of the
s.p. wavefunctions $\phi_\alpha(t)$ according to TDLDA, can be summarized formally as
\begin{subequations}
\label{eq:KSpropag}
\begin{eqnarray}
  |\phi_\alpha(t)\rangle
  &=&
  \hat{U}(t,t')|\phi_\alpha(t')\rangle
  \;,
\label{eq:KSpropag2}
\\
  \hat{U}(t,t')
  &=&
  \hat{\mathcal{T}}\exp\left(-\mathrm{i}
  \int_t^{t'}\hat{h}(t'')dt''\right)
  \;,
\label{eq:timeevol}
\\
  \hat{h}(t)
  &=&
  \frac{\hat{p}^2}{2m}
  +
  U_\mathrm{KS}[\rho(\mathbf{r},t)]
  \;,
\label{eq:UKS}
\end{eqnarray}
\end{subequations}
where $\hat{U}(t,t')$ is the unitary one-body time-evolution operator
with $\hat{\mathcal{T}}$ therein being the time-ordering operator,
$\hat{h}$ is the Kohn-Sham mean-field operator, and $U_\mathrm{KS}$
the (density dependent) actual Kohn-Sham potential \cite{Dre90}.

\subsection{Brief review on RTA}

Mere TDLDA is formulated in terms of a set of occupied single-particle
(s.p.) wavefunctions $\{|\phi_\alpha(t)\rangle,\alpha=1...N\}$
propagating according to eq. (\ref{eq:KSpropag}).  So far, TDLDA deals
with pure Slater states. Dissipation leads inevitably to mixed states.
These can be described compactly by the one-body density operator,
which reads, in natural orbitals representation,
\begin{equation}
  \hat{\rho}
  =
  \sum_{\alpha=1}^\Omega|\phi_\alpha\rangle W_\alpha\langle\phi_\alpha|
\label{eq:rhodiag}
\end{equation}
where $\Omega$ is the size of the configuration space,  which is 
significantly larger than the actual electron number $N$.  The weights
$W_\alpha$ represent the occupation probability for s.p.  state
$|\phi_\alpha\rangle$.  The pure mean-field propagation leaves the
occupation weights $W_\alpha$ unchanged and propagates only the
s.p. states,  such that
$\hat{\rho}(t)=\sum_{\alpha=1}^\Omega |\phi_\alpha(t)\rangle
W_\alpha\langle\phi_\alpha(t)| =
\hat{U}(t,0)\hat{\rho}(0)\hat{U}^{-1}(t,0) $ 
with $\hat{U}$ according to Eq. (\ref{eq:timeevol}).

Dynamical correlations generate time-evolution changes also for the
occupation weights. The RTA describes this in terms of the
density-matrix equation \cite{Rei15d}
\begin{subequations}
\begin{equation}
  \partial_t\hat{\rho}
  +
  \mathrm{i}\big[\hat{h}[\varrho],\hat{\rho}\big]
  =
  \frac{1}{\tau_\mathrm{relax}}
  \left(\hat{\rho}-\hat{\rho}_\mathrm{eq}[\varrho,\mathbf{j},E]\right)
  \;,
\label{eq:RTAbasic}
\end{equation}
where $\hat{h}[\varrho]$ is the Kohn-Sham Hamiltonian
Eq. (\ref{eq:UKS}) in LDA (with ADSIC) depending on the actual local
density distribution
$
\varrho(\mathbf{r},t)=\sum_\alpha{W}_\alpha|\phi_\alpha(\mathbf{r},t)|^2.
$
The right-hand-side stands for the collision term in RTA.  It
describes relaxation towards the local-instantaneous equilibrium state
$\hat{\rho}_\mathrm{eq}[\varrho,\mathbf{j},E]$ for given local density
$\varrho$, current distribution $\mathbf{j}$ and total energy $E$.
The relaxation time $\tau_\mathrm{relax}$ is estimated in
semi-classical Fermi liquid theory. For the metal clusters serving as
test examples in the following, it becomes
\begin{equation}
  \frac{\hbar}{\tau_\mathrm{relax}}
  =
  {0.40}\frac{\sigma_{ee}}{r_s^2}\frac{{E}^*_\mathrm{intr}}{N}
  \quad,
\label{eq:relaxtime}
\end{equation}
\end{subequations}
where $E^*_\mathrm{intr}$ is the intrinsic (thermal) energy of the
system, $N$ the actual number of electrons, $\sigma_{ee}$ the
in-medium electron-electron cross section, and
$r_s=(3/(4\pi\overline{\varrho}))^{2/3}$ is the Wigner-Seitz radius of
the electron cloud \cite{Rei15d}.  It employs an average density
$\overline{\varrho}$ because $\tau_\mathrm{relax}$ is a global
parameter. This approximation is legitimate for metallic systems where
the electron density is rather homogeneous remaining generally close
to the average.  Note that the in-medium cross section $\sigma_{ee}$
also depends on this average density through the density dependence
screening effects.  The actual $\sigma_{ee}$ is taken from the careful
evaluation of \cite{Koe08a,Koe12a} computing electron screening for
homogeneous electron matter in Thomas-Fermi approximation. This yields
$\sigma_\mathrm{ee}=6.5$ a$_0^2$ for the case of Na clusters for
$r_s\approx 3.7$ a$_0$. These are the values which are used throughout
this paper.

The most demanding task is to determine the instantaneous equilibrium
density-operator $\hat{\rho}_\mathrm{eq}[\varrho,\mathbf{j},E]$ in the
RTA equation Eq. (\ref{eq:RTAbasic}). It is the thermal mean-field state
of minimum energy under the constraints of given local density
$\varrho(\mathbf{r})$, local current $\mathbf{j}(\mathbf{r})$, and
total energy $E$.  For the wavefunctions we use the density
constrained mean-field (DCMF) techniques as developed in \cite{Cus85a},
extended to account also for the constraint on current
$\mathbf{j}(\mathbf{r})$. The s.p. states are given occupations
weights $W^\mathrm{(eq)}_\alpha$ according to thermal equilibrium. The
temperature $T$ is tuned to reproduce the desired total energy $E$.
For details of this cumbersome procedure see \cite{Rei15d}.

Once this DCMF step is under control, the RTA scheme is
straightforward. The collision term in Eq. (\ref{eq:RTAbasic}) is
evaluated at time intervals $\Delta t$, typically 0.25 fs and for high
laser frequencies somewhat shorter. In between, the s.p. wavefunctions
in the one-body density are propagated by mean-field evolution
Eq. (\ref{eq:timeevol}). Once one time span $\Delta t$ is completed, we
stay at time $t\!+\!\Delta t$ and dispose of a mean-field propagated,
preliminary one-body density $\tilde{\rho}$ and we evaluate the
collision term.  First, the actual $\varrho$, $\mathbf{j}$, and $E$
are computed. These are used to determine the local-instantaneous
equilibrium state $\hat{\rho}_\mathrm{eq}$.  This is used to step to
the new one-body density $\rho(t\!+\!\Delta{t})=\tilde{\rho}+(\Delta
t/\tau_\mathrm{relax})
\big(\hat{\rho}-\hat{\rho}_\mathrm{eq}[\varrho,\mathbf{j},E]\big)$.
In a final clean-up, this new state $\rho(t\!+\!\Delta t)$ is mapped
into natural orbitals representation Eq. (\ref{eq:rhodiag}), thus delivering
the new s.p. wavefunctions $\varphi_\alpha(t\!+\!\Delta t)$ and
occupation weights $W_\alpha(t\!+\!\Delta t)$ from which on the next
step is performed. For more details see again \cite{Rei15d}.

\section{Energies as key observables}
\label{sec:energies}

In our  previous paper on RTA, we have concentrated on thermalization
processes, in particular on relaxation times \cite{Rei15d}. Here, we
are going to employ RTA to the energy balance in metal clusters
excited by strong laser fields. The key observables are the various
contributions to the energy which we will introduce in this section.
The expressions assume tacitly a numerical representation of
wavefunctions and fields on a spatial grid in a finite box with
absorbing boundaries. Particularly the boundaries require some care as we
will see.

The basic question we aim to analyze here is how the energy absorbed
by the laser is "used" by the cluster and redistributed into various
well identified components. The key starting quantity will thus be the
energy absorbed by the laser which we denote by
$E^{\mbox{}}_\mathrm{abs}$.  The basic energy branching channels
of the cluster consist in electron emission and intrinsic heating
\cite{Rei15d} and we thus have to analyze both these components
separately. Electron emission corresponds to charge loss
associated with energy loss because the emitted electrons carry
some energy outwards. We denote this energy by
$E^{\mbox{}}_\mathrm{ch,loss}$.  But electron emission also affects
the cluster itself, net cluster charge leading to an associated
change in potential energy $E^{\mbox{}}_\mathrm{ch,pot}$. The
remaining energy dekivered by the laser is shared between collective
motion of electron leading to collective kinetic energy
$E^{\mbox{}}_\mathrm{coll}$ and "intrinsic" excitation energy of the
electron cloud itself consisting out of a kinetic
$E^{\mbox{}}_\mathrm{intr,kin}$ and a potential
$E^{\mbox{}}_\mathrm{intr,pot}$ component.  All terms, of course, sum
up to $E^{\mbox{}}_\mathrm{abs}$:
\begin{equation}
  E^{\mbox{}}_\mathrm{abs}
  =
  E^{\mbox{}}_\mathrm{ch,loss}\!+\!E^{\mbox{}}_\mathrm{ch,pot}
  +
  E^{\mbox{}}_\mathrm{intr,kin}\!+\!E^{\mbox{}}_\mathrm{intr,pot}
  \!+\!E^{\mbox{}}_\mathrm{coll}
  \quad,
\label{eq:compl}
\end{equation}
Let us now specify these various contributions in more detail. This
implies that we also detail small components related to the treatment
of absorbing boundaries conditions and which have to be properly
accounted for in the energy balance. Moreover, we introduce as
auxiliary quantity the actual total energy $E(t)$ of the system which
is a crucial input for the RTA step.  The various energy components
are thus computed as follows:
\begin{enumerate}
 \item {\bf $\boldsymbol{E_\mathrm{abs}}=$  Energy absorbed from the laser field:}
  \begin{equation}
    E_\mathrm{abs}
    =
    \int_0^t dt'\,\int d^3r\mathbf{E}_0(t')\cdot\mathbf{j}(\mathbf{r},t')
    -E_\mathrm{abs}^\mathrm{(mask)}
  \label{eq:Eabs}
  \end{equation}
    where $E_\mathrm{abs}^\mathrm{(mask)}$ is a correction for the
    particle loss at the absorbing bounds (for details see appendix
    \ref{app:boundcorr}).  
\\
 \item {\bf $\boldsymbol{E(t)=}$ total energy:}\label{it:totalE}
  \begin{eqnarray}
    E(t)
    &=&
    E^*_\mathrm{TDLDA}(t)
    +
    E_\mathrm{pot,bc}
    \;,
  \label{eq:Et}
  \\
    E_\mathrm{pot,bc}
   &=&
    \int_0^t dt'\int d^3r\,(1-\mathcal{M}^2)U_\mathrm{KS}\rho(\mathbf{r},t')
    \;.
  \end{eqnarray}
  Thereby $E^*_\mathrm{TDLDA}(t)=E_\mathrm{TDLDA}(t)-E_\mathrm{g.s.}$
  is the energy $E_\mathrm{TDLDA}(t)$ computed with the given
  LDA+ADSIC functional taken relative to the static ground state
  energy $E_\mathrm{g.s.}$.  The $E_\mathrm{pot,bc}$ is a correction
  for the small amount of binding energy carried in the absorbed
  electrons, an artifact which arises due to finite numerical
  boxes. Altogether, $E(t)$ accounts for the energy left within the
  simulation box as result of energy absorption from the laser and
  energy loss through ionization.  \\
 \item {\bf $\boldsymbol{E_\mathrm{ch,loss}=}$ energy loss by
   electron emission}:\\
   \begin{equation}
     E_\mathrm{ch,loss}=E_\mathrm{abs}-E(t)
   \end{equation}
  It represents the kinetic energy carried away by the emitted
  electrons.
\\
 \item {\bf $\boldsymbol{E_\mathrm{ch,pot}(Q)=}$ charging energy}:
   \begin{equation}
     E_\mathrm{ch,pot}(Q)
     =
     E_\mathrm{g.s.}(Q)-E_\mathrm{g.s.,initial}
     -E_\mathrm{pot,bc}
   \label{eq:charge}
   \end{equation}
   where $E_\mathrm{g.s.}(Q)$ is the ground state energy
   (i.e. temperature $T=0$) for a given
   charge state $Q$ and $E_\mathrm{g.s.,initial}=E(t\!=\!0)$ the
   initial ground state energy. For compensation of definition
   (\ref{eq:Et}), it is augmented by the correction for lost potential energy.
   The $E_\mathrm{ch,pot}(Q)$ accounts for the excitation energy
   invested for charging the cluster. 
\\
 \item {\bf ${\boldsymbol E_\mathrm{intr,kin}=}$ intrinsic kinetic energy :}\\
  \begin{equation}
    E_\mathrm{intr,kin}
    =
    E_\mathrm{TDLDA}(t)-E_\mathrm{DCMF}(\varrho,\mathbf{j},T\!=\!0)
  \label{eq:EintrDCMF}
  \end{equation}
  where $E_\mathrm{TDLDA}(t)$ is the actual LDA+ADSIC energy and
  $E_\mathrm{DCMF}(\varrho,\mathrm{j},T\!=\!0)$ the DCMF energy at $T=0$
  (= ground state for fixed $\varrho$ and $\mathbf{j}$). The computation
  is simplified by exploiting the fact that $\varrho$ and $\mathbf{j}$
  remain frozen in DCMF and  thus also the Kohn-Sham
  potential. This allows to take the difference of the sums of
  s.p. kinetic energies between the two configurations.
\\
 \item {\bf ${\boldsymbol E_\mathrm{intr,pot}=}$ intrinsic potential energy :}\\
  \begin{equation}
    E_\mathrm{intr,pot}
    =
    E_\mathrm{DCMF}(\rho,\mathbf{j}\!=\!0,T\!=\!0)-E_\mathrm{g.s.}(Q)
    \quad.
  \label{eq:intpot}
  \end{equation}
  This is the ``potential'' energy stored in the constraint on given
  $\rho$ \& $\mathbf{j}$ at $T=0$.
\\
 \item {\bf $\boldsymbol{E_\mathrm{coll}=}$ collective flow energy:}\\
   \begin{equation}
     E_\mathrm{coll}
     =
     \int d^3r\,\frac{\mathbf{j}^2(\mathbf{r})}{2m\rho(\mathbf{r})}
   \label{eq:Ecoll}
   \end{equation}
   This is the kinetic energy which is contained in the average
   momentum distribution $\mathbf{j}(\mathbf{r})$. It is to be
     noted that
     $E_\mathrm{coll}=E_\mathrm{DCMF}(\rho,\mathbf{j},T\!=\!0)-
     E_\mathrm{DCMF}(\rho,\mathbf{j}\!=\!0,T\!=\!0)$. This shows that
     $E_\mathrm{coll}$ is part of the intrinsic energy.
\end{enumerate}

For the balance plots below, we consider also the relative
contributions $E_\mathrm{ch,loss}/E_\mathrm{abs}$,
$E_\mathrm{ch,pot}/E_\mathrm{abs}$,
$E_\mathrm{intr,kin}/E_\mathrm{abs}$,\\ $E_\mathrm{intr,pot}/E_\mathrm{abs}$,
and $E_\mathrm{coll}/E_\mathrm{abs}$ adding up to one. Moreover,
  we use the completeness Eq. (\ref{eq:compl}) to deduce
  $E_\mathrm{intr,pot}$ from the other energies. This saves another
costly DCMF evaluation for
$E_\mathrm{DCMF}(\varrho,\mathbf{j}\!=\!0,T\!=\!0)$ in the definition
Eq. (\ref{eq:intpot}).

Finally, we mention that the evaluation of the
intrinsic kinetic energy Eq. (\ref{eq:EintrDCMF}) had been used in the
past often with a semi-classical estimate \cite{Cal00}, for details
see appendix \ref{app:semicl}. This is much simpler to evaluate, but
not precise enough for the present purposes. Moreover, we need the
expensive DCMF state anyway and so get the correct quantum mechanical
value Eq. (\ref{eq:EintrDCMF}) for free.

\section{Results and discussion}
\label{sec:results}

In the previous RTA paper \cite{Rei15d}, we had briefly looked at
dissipation effects as function of laser frequency for constant
intensity and found that dissipation is strong if the laser is in
resonance with a system mode and weak otherwise. This is a trivial
result in view of Eq. (\ref{eq:relaxtime}): The relaxation rate
increases with excitation energy and excitation energy is large at
resonance. In order to eliminate this trivial trend, we consider here
variation of laser parameters for fixed absorbed energy
$E_\mathrm{abs}$ tuning the intensity such that the wanted value for
$E_\mathrm{abs}$ is maintained. We calibrate the laser parameters this
way for the case of pure TDLDA and use the same parameters then also
for RTA. The resulting $E_\mathrm{abs}$ is in most situations the
same. A difference in $E_\mathrm{abs}$ between RTA and TDLDA, if it
occurs, is then already a message.

One of the interesting topics related to energy balance is the
question of appearance size, the limit of fission/fragmentation
stability of a metal cluster for a given charge state
\cite{Cha95,Nae97}.  It is the lower the more gentle one can arrange
ionization.  The systematics of energy balance will tell us how to
ionize most gently or, in reverse, to heat most efficiently.

\subsection{Typical time evolution of energies}
\label{sec:timevol}

\begin{figure}
\centerline{\includegraphics[width=0.93\linewidth]{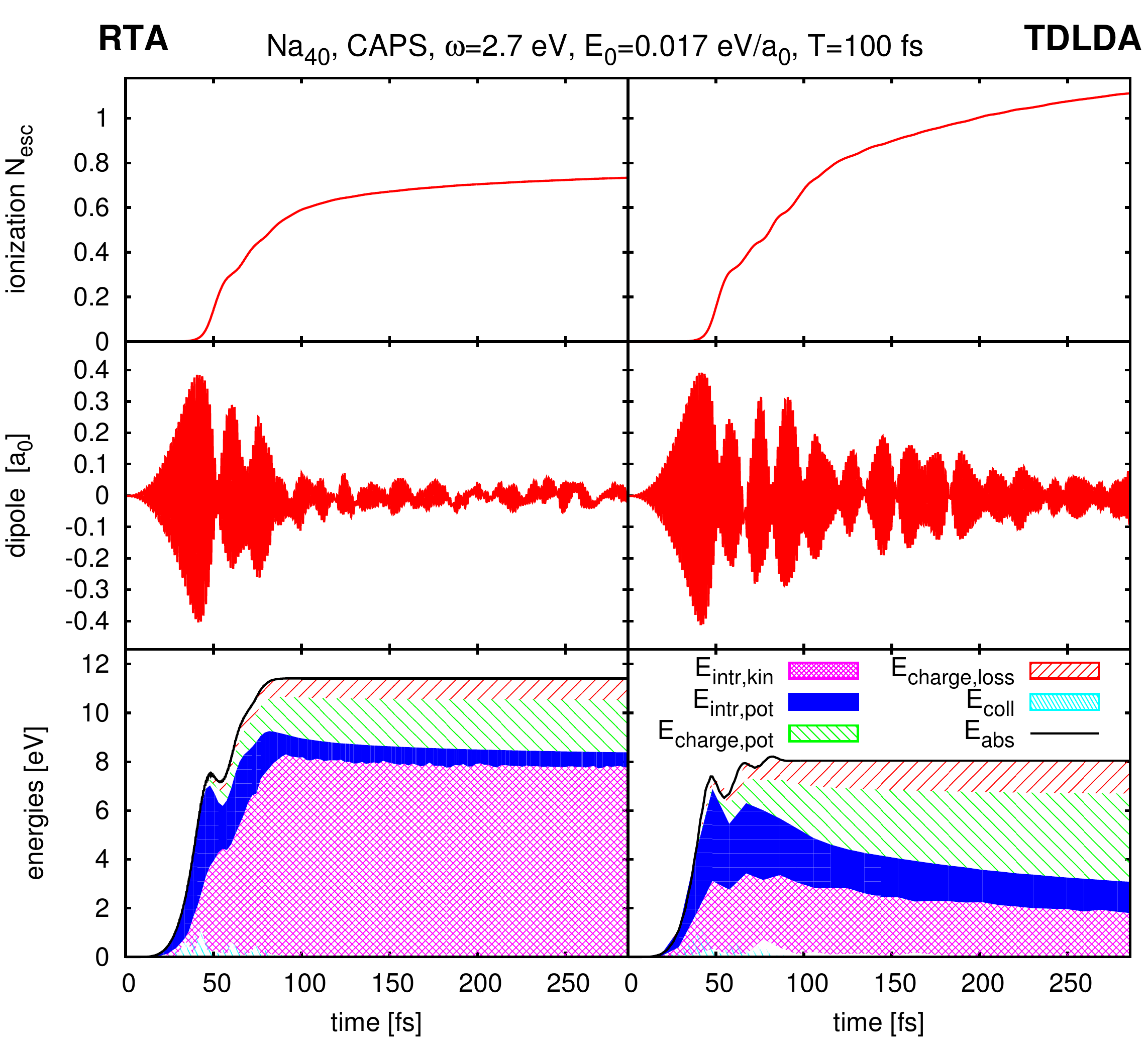}}
\caption{\label{fig:na40_o20_e00122_balance} Time evolution of
  ionization (upper panels), dipole moment (middle panels), and
  energies (lower panels) for the case of Na$_{40}$ in CAPS excited by
  a laser with frequency $\omega=2.7$ eV, total pulse length
  $T_\mathrm{pulse}=100$ fs, and intensity $I=1.3\,10^{10}$ W/cm$^2$.
  Left panels show results from RTA and right panels from TDLDA.  The
  lower panels show the total absorbed energy (black line) ad the four
  different contributions stacked one above the other.}
\end{figure}
The lower panel of figure \ref{fig:na40_o20_e00122_balance} shows the
time evolution of the five contributions Eq. (\ref{eq:compl}) to the
energy stacked in a balance manner. Each colored band represents the
contribution indicated in the key to the right side of the
panels. Upper and middle panels show as complementing information
dipole moment and ionization.
The case $\omega=2.7$ eV shown in Figure
\ref{fig:na40_o20_e00122_balance} corresponds to a resonant excitation
at the Mie plasmon frequency.  We see this from the time evolution of
ionization $N_\mathrm{esc}\equiv Q$ and dipole. The TDLDA result
(right panels) shows ongoing dipole oscillations and, connected with
that, ionization carries on long after the laser pulse has
terminated. However, the RTA ionization (upper left panel) turns
gently to a constant $N_\mathrm{esc}$.  This is achieved by the
dissipation in RTA which damps the dipole signal.  This highly
resonant case reveals a marked qualitative difference between TDLDA
and RTA. We thus see that long-time TDLDA simulations have to be taken
with care because they overestimate the long-lasting reverberations of
the dipole.

The difference in ionization  also shows up as a difference in the
energy loss by ionization (green and blue areas) such that eventually
TDLDA produces relatively less intrinsic excitation energy in than
RTA.

The lower panels of figure \ref{fig:na40_o20_e00122_balance}  also show
the collective kinetic energy Eq. (\ref{eq:Ecoll}). It plays a role in the
initial stages of excitation. The reason is that the dipole field of
the laser couples to the collective dipole operator thus depositing
its energy first in collective dipole flow. However, the large
spectral fragmentation of the dipole mode (Landau damping)
\cite{Rei96b,Bab97} spreads the collective energy very quickly over
the dipole spectrum. The large fragmentation width of the actual test
case Na$_{40}$ produces a relaxation time below 1 fs for this Landau
damping and this relaxation takes place already at mean field level.
As a consequence, collective kinetic energy becomes negligible soon
after the laser pulse is extinguished. We will ignore it in the
following analysis evaluated at late stages of the cluster dynamics.

It is remarkable that RTA allows to absorb much more energy
$E_\mathrm{abs}$ from the laser, although exactly the same pulse is
used in both cases. This is a particular feature of resonant
excitation related to Rabi oscillations \cite{Lou09aB}.  The external
field quickly induces dipole oscillations of the electron cloud.  This
dipole excitation, once sufficiently large, leads to stimulated
emission and so reduces excitation. 
This can be seen from oscillations of $E_\mathrm{abs}$ where
phases of energy absorption are interrupted by phases of energy loss
back to the laser field. 
Now in RTA, dissipation serves as
a competing de-excitation channel which reduces stimulated emission
and so paves the way to more stimulated absorption. This mechanism is
less important off resonance where we observe generally less
differences between RTA and TDLDA as we will see in the upper panel of
figure \ref{fig:na40_E60_observables}.

\subsection{Trends with laser frequency}

The main intention of the study is to figure out trends with laser
parameters. To this end, we simulate each case for a time of 300 fs
and collect the results at this final time. This is a safe procedure
for the majority of non-resonant cases. It is incomplete for resonant
excitation, at least with TDLDA. In the latter case we have to 
keep in mind that the contribution of emission is somewhat
underestimated and that of intrinsic energy overestimated. The major
trends remain, nonetheless, the same.

\begin{figure}
\centerline{\includegraphics[width=\linewidth]{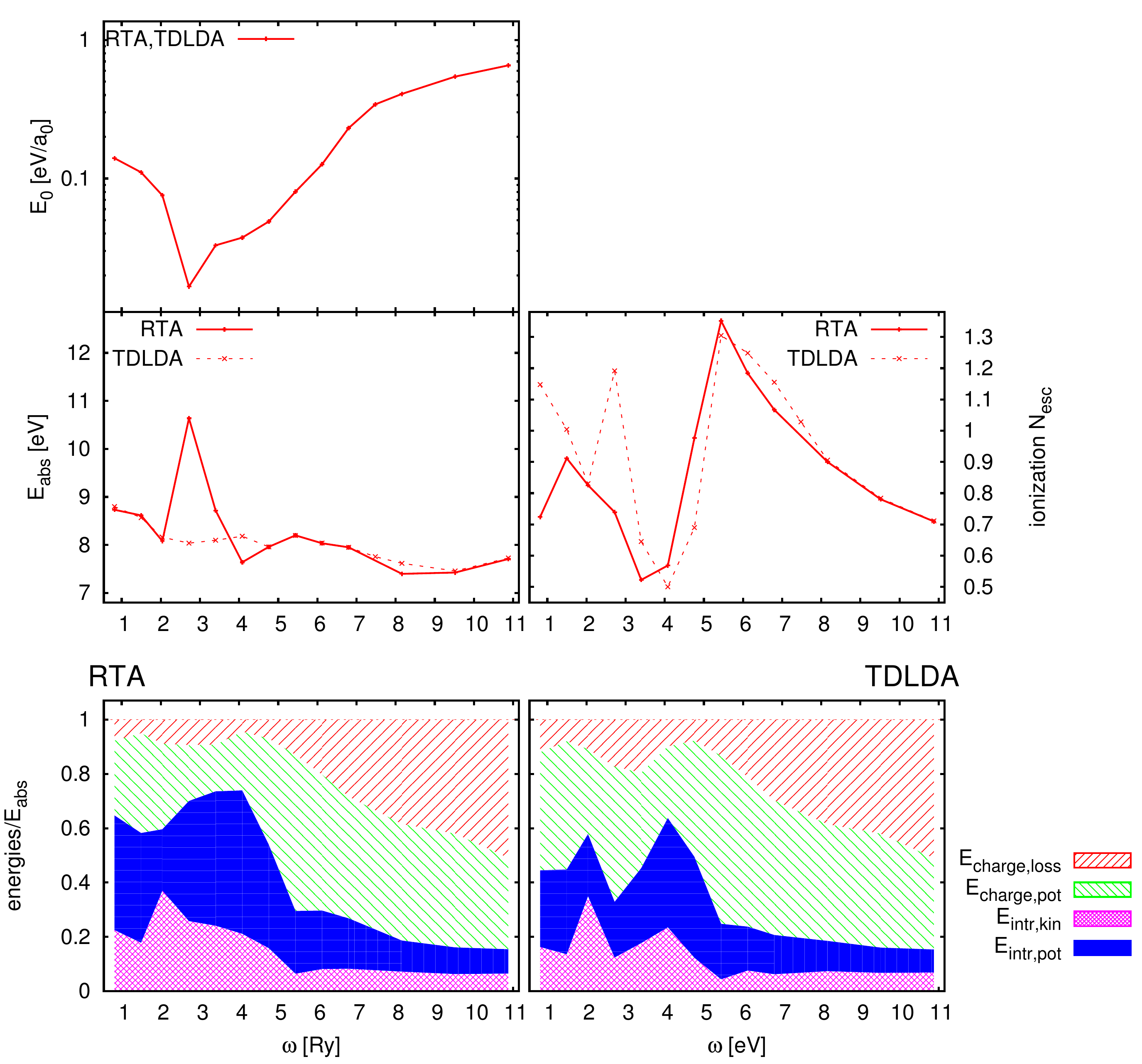}}
\caption{\label{fig:na40_E60_observables}
Various observables from RTA (full lines) and TDLDA (dashed lines)
 evaluated at final time of the simulations at 300 fs.
Pulse length was 
$T_\mathrm{pulse}=100$ fs throughout. Intensity has been
tuned such that $E_\mathrm{abs}\approx 8.2$ eV for TDLDA. 
Upper left: field strength $E_0$ ($\propto \sqrt{I}$).
Middle left: total absorbed energy $E_\mathrm{abs}$.
Middle right: ionization $N_\mathrm{esc}$.
Lower: Balance of relative energies (energy contributions
divided by total absorbed energy $E_\mathrm{abs}$). 
Left panel for RTA and right one for TDLDA.
}
\end{figure}
Figure \ref{fig:na40_E60_observables} shows the energy contributions
and other observables as function of laser frequency $\omega$. The
laser intensity is tuned for each frequency such that the absorbed
energy is about the same, namely $E_\mathrm{abs}\approx 8.2$ eV, for
TDLDA. The same field strength is then used also for RTA and the
emerging $E_\mathrm{abs}$ may then be different. This is indeed seen
in the left middle panel where just near the Mie plasmon resonance
($\approx 2.7$ eV) RTA absorbs much more energy, as was discussed
already in connection with figure \ref{fig:na40_o20_e00122_balance}.

The upper left panel of Figure \ref{fig:na40_E60_observables} shows
the field strength $E_0$. The Mie plasmon resonance is visible as
marked dip at $\omega=2.7$ eV because resonance means that more response
is achieved with less impact.  The steady growth of $E_0$ for larger
frequencies complies with the Keldysh formula where the effective
field strength shrinks $\propto\omega^{-2}$ \cite{Kel65}.

The middle right panel shows ionization $N_\mathrm{esc}$. At lower
frequencies, RTA suppresses emission significantly. Obviously, more of
the absorbed energy is turned to intrinsic excitation
(thermalization).  Quite different is the behavior at high frequencies
above ionization potential (IP) in which case TDLDA and RTA deliver
almost the same ionization.

The lower panels disentangle the absorbed energy into its four
relevant contributions (\ref{eq:compl}).  Again, we see that TDLDA and
RTA differ most at the side of lower energies, particularly near the
Mie plasmon resonance. There is practically no difference from
$\omega\approx 6.1$ eV on. This $\omega=6.1$ eV is a very prominent
point. It is just the frequency from which on all occupied valence
electrons of Na$_{40}$ can be emitted by a one-photon process. The IP
at 3.5 eV Ry sets the frequency where the HOMO can be removed by one
photon. The region 3.5-6.1 eV covers the transition from the onset of
one-photon processes for the least bound state to an ``all inclusive''
one-photon ionization. And we see, indeed, how TDLDA and RTA results
come stepwise closer to each other in this region.

The lower panels of figure \ref{fig:na40_E60_observables} shows the
results in form of energy balance where the filled areas visualize a
given contribution, as indicated.  Blue and green areas together show
the amount of energy spent for ionization while purple and yellow
together illustrate the part of the intrinsic energy.  The balance
plot makes the trends of intrinsic energy immediately visible. Its
fraction is largest around resonance and smallest above the
point of ``all one photon'' ionization near $\omega=6.1$ eV. 
This
trend holds for RTA as well as for TDLDA.  What differs are the actual
fractions of intrinsic energy, generally being somewhat larger for
RTA. But the fractions are not so dramatically different as one may
have expected from the plot of energies as such in figure
\ref{fig:na40_o20_e00122_balance}. Division by $E_\mathrm{abs}$ and
the often larger $E_\mathrm{abs}$ in RTA reduces the effect for the
fractions of energy.

Already at this point, we can give a first answer to the question of
how to ionize most gently or to heat most efficiently. Least intrinsic
energy relative to most electron output is achieved near the point
from which on all electrons can be removed by one photon which is
$6.1$ eV in the present case. Most heating is obtained below,
particularly near resonance or for very low frequencies.

\subsection{Trends with pulse length $T_\mathrm{pulse}$}

\begin{figure}
\centerline{\includegraphics[width=\linewidth]{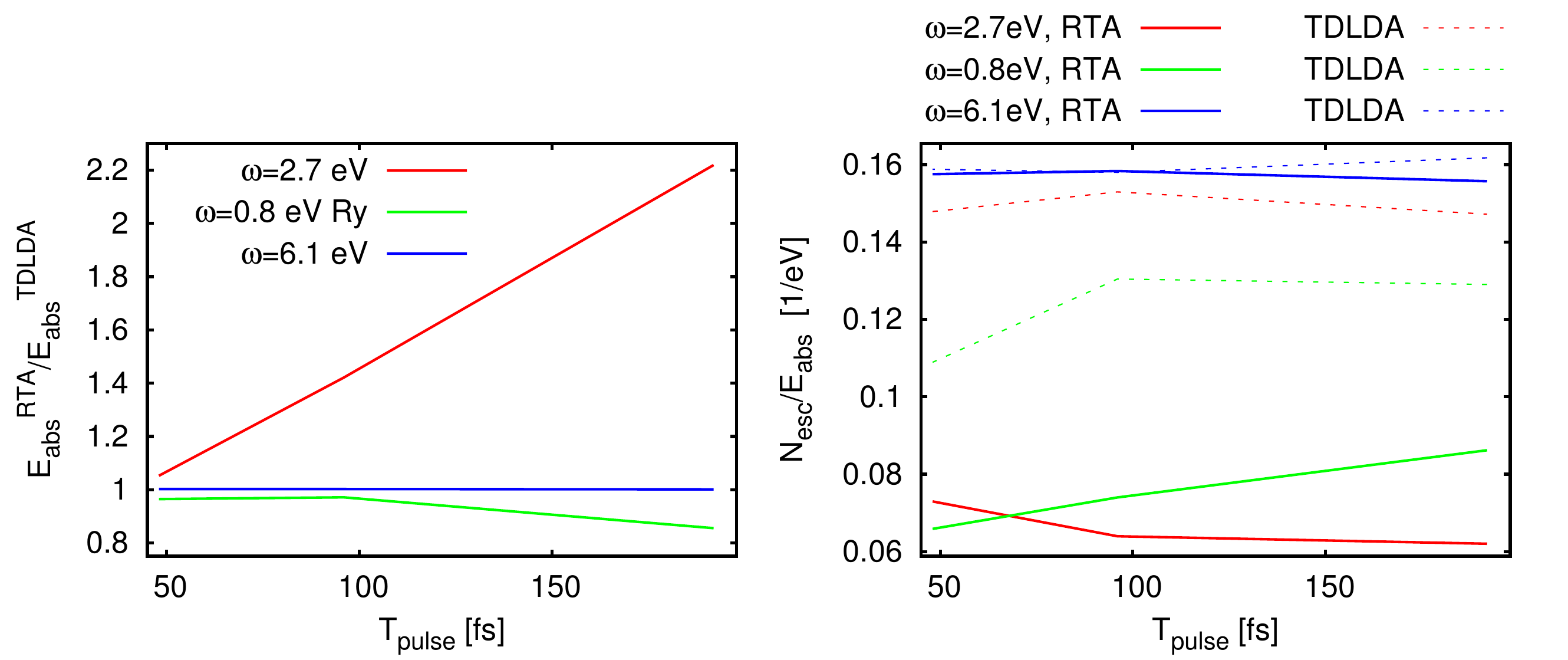}}
\centerline{\includegraphics[width=\linewidth]{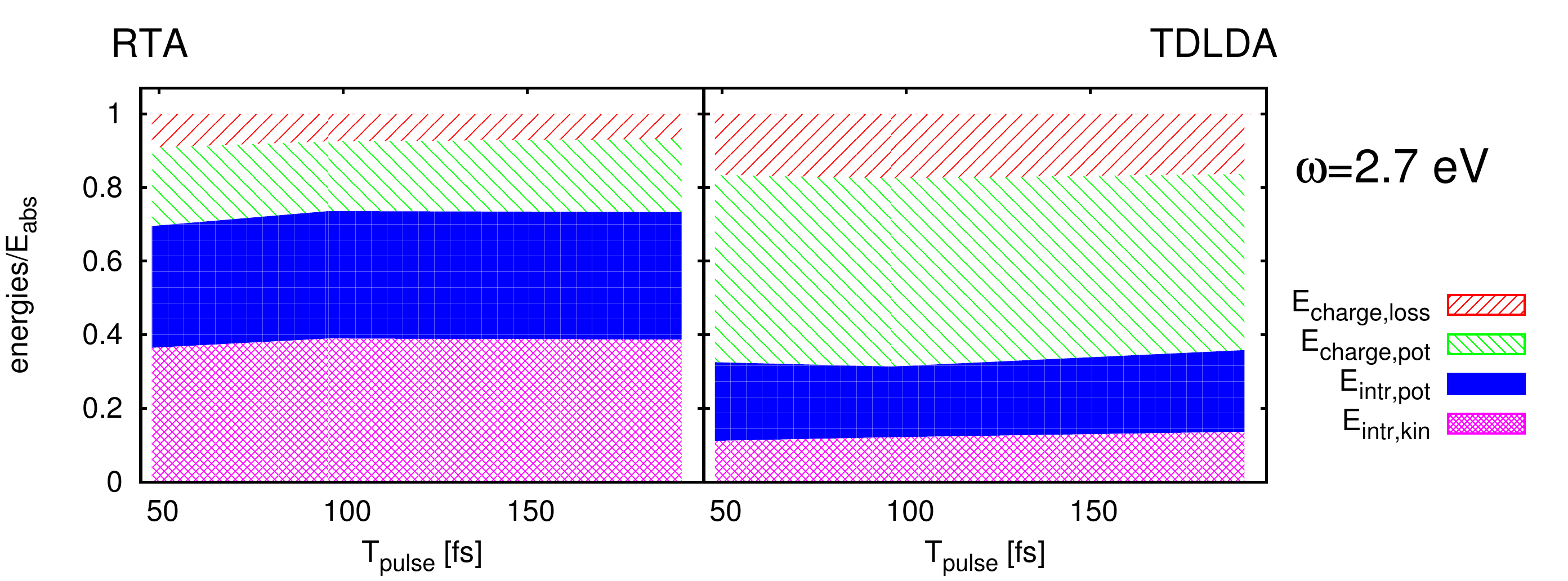}}
\caption{\label{fig:na40_varytime_balance} Lower panel: Energy balance
  as function of pulse length $T_\mathrm{pulse}$ for Na$_{40}$ excited
  by a laser with three frequency $\omega=2.7$ eV and intensity tuned to
  $E_\mathrm{abs}^\mathrm{(TDLDA)}\approx 8.2$ eV.  Left upper panel: The
  ratio
  $E_\mathrm{abs}^\mathrm{(RTA)}/E_\mathrm{abs}^\mathrm{(TDLDA)}$ for
  three different frequencies as indicated, resonant $\omega=2.7$ Ry
  and off-resonant $\omega=0.8,6.1$ eV.  Right upper panel: Ratio
  $N_\mathrm{esc}/E_\mathrm{abs}$ of emitted electrons per absorbed
  energy for the three frequencies as in the left panel and separately
  for RTA (full lines) as well as TDLDA (dashed lines).  }
\end{figure}
Figure \ref{fig:na40_varytime_balance} shows the effect of laser pulse
length $T_\mathrm{pulse}$. The lower panels show the energy balance.
as function of $T_\mathrm{pulse}$ for the resonant case $\omega=2.7$ eV.
The trends with $T_\mathrm{pulse}$ are extremely weak, even for the
most sensitive case of resonant excitation. They are equally weak for
other frequencies. Thus these are not shown.

One interesting aspect pops up, again, concerning the amount of absorbed
energy.  This is illustrated in the upper panel of figure 
\ref{fig:na40_varytime_balance} showing the ratio from RTA to TDLDA,
$E_\mathrm{abs}^\mathrm{(RTA)}/E_\mathrm{abs}^\mathrm{(TDLDA)}$, for
three frequencies standing for the three typical regions, very low
frequency (0.8 eV), resonance (2.7 eV), and above threshold for direct
ionization of all shells (6.1 eV). The energy ratio increases
dramatically with pulse length in the resonant case $\omega=2.7$ eV.
Although the partition of energies changes very little, the total
output becomes much larger with RTA for long pulses.  This happens
because dissipation steadily removes energy from the coherent dipole
oscillations thus keeping the door open for ongoing energy absorption
while in TDLDA energy loss by stimulated emission limits energy
take-up, see the discussion in section \ref{sec:timevol}.  For
off-resonant cases, the ratio
$E_\mathrm{abs}^\mathrm{(RTA)}/E_\mathrm{abs}^\mathrm{(TDLDA)}$ stays
close to one as can be seen here for low frequency $\omega=0.8$ eV and
for high frequency 6.1 eV.

\subsection{Trends with field strength (laser intensity)}

\begin{figure}[h]
\centerline{\includegraphics[width=\linewidth]{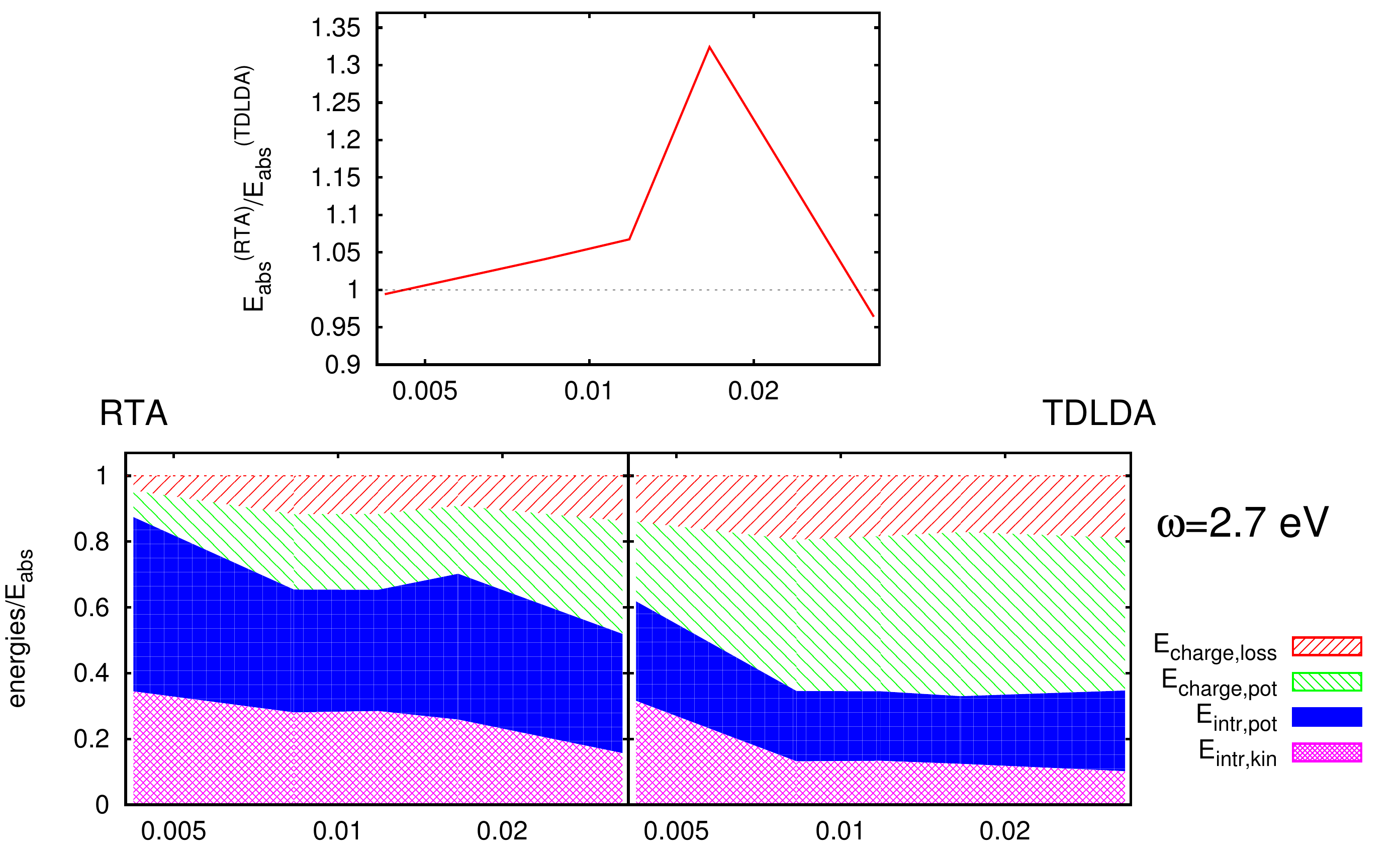}}
\centerline{\includegraphics[width=\linewidth]{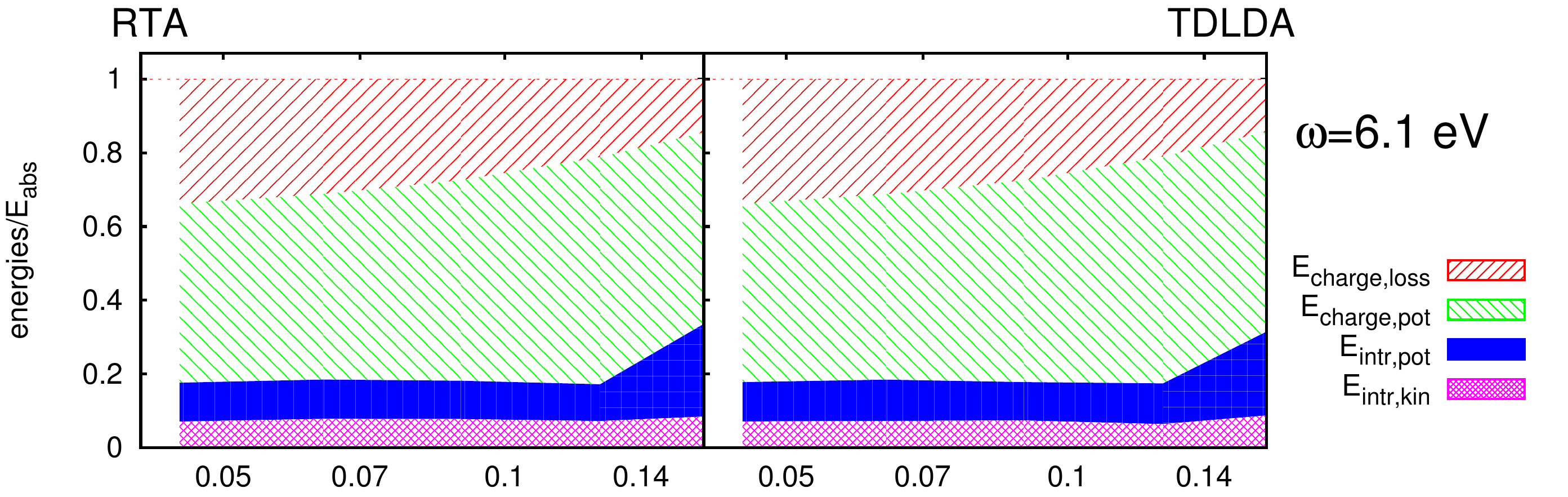}}
\centerline{\includegraphics[width=\linewidth]{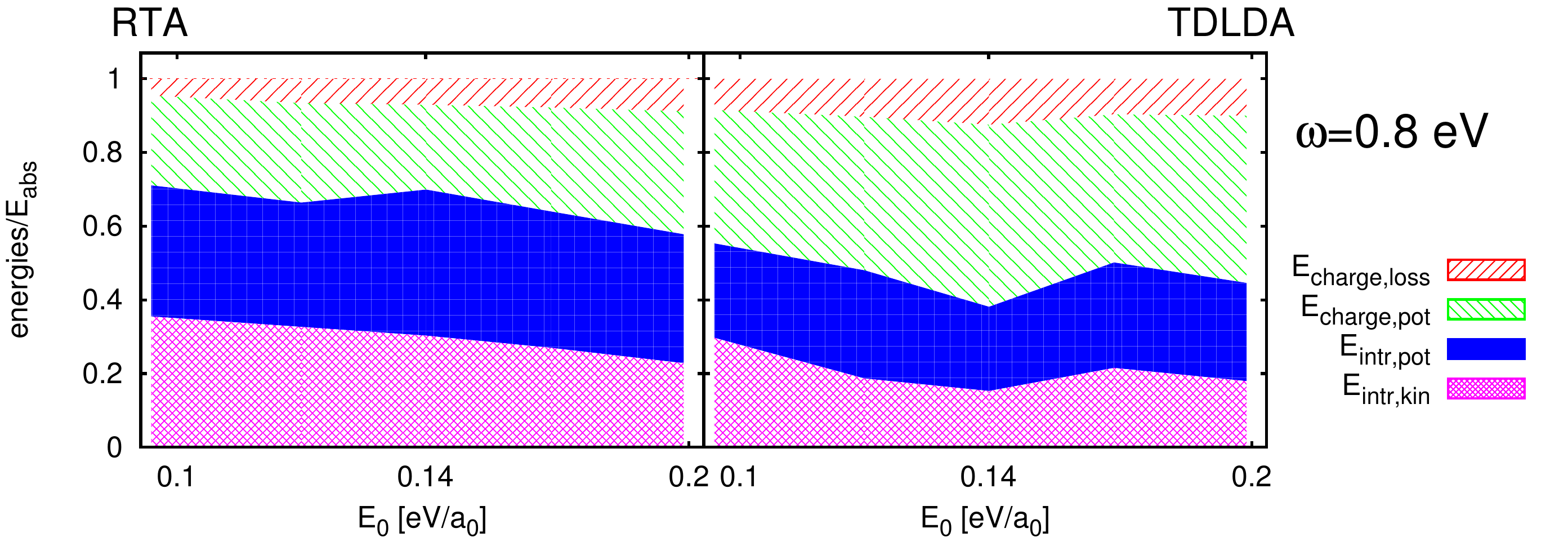}}
\caption{\label{fig:na40_varystrength_balance} Lower three panels:
  Energy balance as function of field strength $E_0$ for Na$_{40}$
  excited by a laser with three different frequencies as indicated and
  pulse length $T_\mathrm{pulse}=100$ fs. Upper panel: Ratio of
  absorbed energy between RTA and TDLDA for the resonant case
  ($\omega_\mathrm{las}=2.7$ eV) as function of field strength. }
\end{figure}
The three lower panels of figure \ref{fig:na40_varystrength_balance}
show the effect of laser field strength $E_0$ for three frequencies,
low $\omega=0.8$ eV, resonant $\omega=2.7$ eV, and high $\omega=6.1$
eV which is on the onset of the one-photon regime for all occupied
s.p. states. For the low frequency and the resonant case, intrinsic
energy shrinks with increasing $E_0$.  The reason is that higher order
photon processes become increasingly important which, in turn,
enhances the contribution from direct (multi-photon) emission leaving
less energy to dissipate. For resonant excitation, we have the
additional effect that  the Mie plasmon frequency is increasing with
increasing $E_0$ because ionization is stronger and enhances the
charge state of the cluster \cite{Rei96b}. Thus the resonance
frequency is running away from the laser frequency which also reduces
dissipation. For the high-frequency case $\omega=6.1$ eV, the
intrinsic energy increases with $E_0$. This is, again, an effect of
ionization which drives the IP up and thus moves large parts of the
s.p. states out of the one-photon regime back to the multi-photon
regime. Differences between the frequencies shrink with increasing
$E_0$ because the fraction of intrinsic energy decreases with $E_0$
for the low and medium frequencies thus approaching the high frequency
case (related to direct emission).  Convergence is better visible
within the given span of $E_0$ for RTA while it requires even larger
$E_0$ for TDLDA.  The effect is plausible because large $E_0$ means
that we come into the field dominated regime where frequencies become
less important and where direct field emission takes over
\cite{Rei99a}.

The upper panel of figure \ref{fig:na40_varystrength_balance} shows
the ratio of absorbed energy
$E_\mathrm{abs}^\mathrm{(RTA)}/E_\mathrm{abs}^\mathrm{(TDLDA)}$ as
function of field strength for the resonant frequency
$\omega_\mathrm{las}=2.7$ eV. This case, unlike the non-resonant
frequencies, shows a peak at a certain field strength. This emerges as
combination from several features seen before. At small field
strengths, there is little energy deposited, thus little dissipation
and RTA does not differ much from TDLDA. More energy becomes absorbed
with increasing field strength which is converted preferably to
intrinsic energy in the resonant case opening subsequently the pathway
to more absorption. This explains the increase from low $E_0$ on
upwards.  For larger amounts of absorbed energy, the enhanced
dissipation broadens the resonance thus reducing resonant response at
peak frequency. This explains the decrease of the ratio for further
increasing field strengths.

\subsection{Impact of cluster charge}
\label{sec:charge}

For the one reference system Na$_{40}$, we have so far studied laser
excitation with extensive exploration of the rich variety of laser
parameters. We are now going to vary the systems under consideration, 
studying clusters of the form  Na$_{40+Q}^{+Q}$ which have $N_{el}=40$ 
electrons and varied charges state $Q$.
It would not make sense to unfold all the laser variations for each
system anew. Thus we take as a means of comparison an instantaneous
dipole boost.
$\varphi_\alpha\rightarrow\exp(-\mathrm{i}\,p_0\,z)\varphi_\alpha$ applied
to all s.p. wavefunctions in the same manner \cite{Cal00,Rei04aB}. The
boost momentum $p_0$ regulates its strength associated with the
initial excitation energy $E_\mathrm{abs}=Np_0^2/(2m)$ which can be
compared with the absorbed energy in the laser case.  The boost
excitation touches all modes of a system at once with some bias on
resonant excitation and it has only one parameter which simplifies
global comparisons between different systems.  We will thus use boost
excitation in this section for variation of cluster charge and in the
next section for cluster size.

There is another subtle problem when varying cluster charge: the ionic
geometry changes with charge state. This can become particularly
pronounced for deformed clusters. Thus we consider variation of charge
for a magic electron number, actually $N_\mathrm{el}=40$. This forces
all systems for any charge state to near spherical geometry. We go one
step further and exclude any geometry effect by using a soft jellium
density for the ionic background \cite{Mon94b,Cal00}.
\begin{figure}[bht]
\centerline{\includegraphics[width=\linewidth]{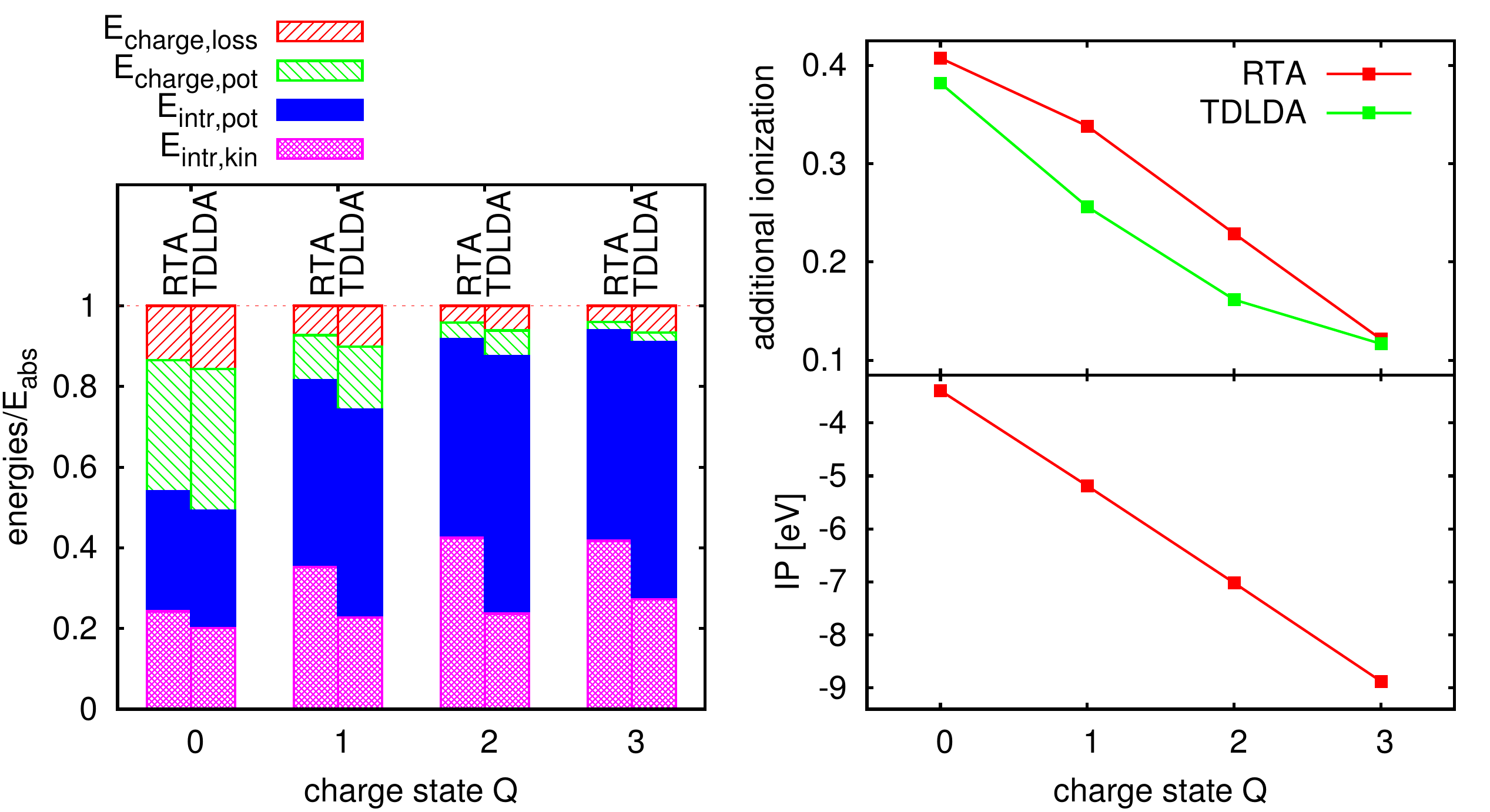}}
\caption{\label{fig:chargescomb-balance} Results for clusters
  Na$_{40+Q}^{+Q}$ which have $N_{el}=40$ electrons and varied
  charges state $Q$.  The ionic structure is approximated by soft
  spherical jellium model with Wigner-Seitz radius $r_S=3.65$ a$_0$
  and surface parameters $\sigma=1$ a$_0$ \cite{Mon94b,Cal00}. All
  clusters are excited initially by an instantaneous boost with boost
  energy $E_\mathrm{abs }=2.7$ eV.  Left panel: Energy balance for Na, 
  plotting RTA and TDLDA side by side.  Lower right: Ionization
  potential (IP).  Upper right: Ionization induced by boost.  }
\end{figure}
The result for charge balance after boost excitation with initial
energy of 2.7 eV is shown in figure \ref {fig:chargescomb-balance}.
We see again the typical pattern: about equal share of intrinsic
kinetic and intrinsic potential, about factor 2 more energy invested
charging the cluster than energy lost by emission, and somewhat more
intrinsic energy in RTA as compared to TDLDA. The new feature here is
that we see a strong trend of the intrinsic energy versus energy loss
by emission. Electron emission decreases with increasing charge state
$Q$ because the IP increases with $Q$ which enhances the cost of
emission. In turn, less energy is exported by emission and invested
into charging energy while more energy is remaining in
the clusters for dissipation into intrinsic excitation energy. The
trend is clear, simple, and monotonous. It will apply equally well
in other systems (with varying IP) and other observations. For example,
laser frequency scans for different charge states will show the same
pattern as function of frequency, but with an increasing offset of
intrinsic energy with increasing charge state.

\subsection{Impact of cluster size}
\label{sec:size}

We have also compared RTA with TDLDA for clusters of different size
considering a series of closed-shell systems Na$_{9}^+$, Na$_{21}^+$,
Na$_{41}^+$, as well as open-shell systems Na$_{15}^+$,
Na$_{33}^+$. This sample allows to explore trends with system size as
well as the effect of shell closures.  As for variation of charge in
the previous section, we avoid a tedious scan of frequencies and other
laser parameters by using simply a boost excitation. Two boost
strengths are considered, $E_\mathrm{boost}/N_\mathrm{el}=0.027$ eV
still in the linear regime and a higher
$E_\mathrm{boost}/N_\mathrm{el}=0.14$ eV. Note that these boost
strength are scaled to system size. This should provide comparable
thermodynamic conditions (e.g. temperatures).

No clear trend with system size could be found. However, at lower
excitation energies, we see a shell effect to the extend that magic
systems gather more thermal energy. This shell effect is going away
for the higher excitations.  It is to be noted that the lower
excitation strength $E_\mathrm{boost}/N_\mathrm{el}=0.027$ eV leads in
all five system to a temperature around 1500 K while the higher
excitation $E_\mathrm{boost}/N_\mathrm{el}=0.14$ eV is associated with
temperature about 3000 K. This matches with observations from shell
structure in Na clusters where the disappearance of shell effects is
located at about 2000 K \cite{Bra93,Mar93}.
The lower excitation strength here is below this critical point and
the higher excitation above.

\subsection{Excitation with by-passing ions}

\begin{figure}
\centerline{\includegraphics[width=\linewidth]{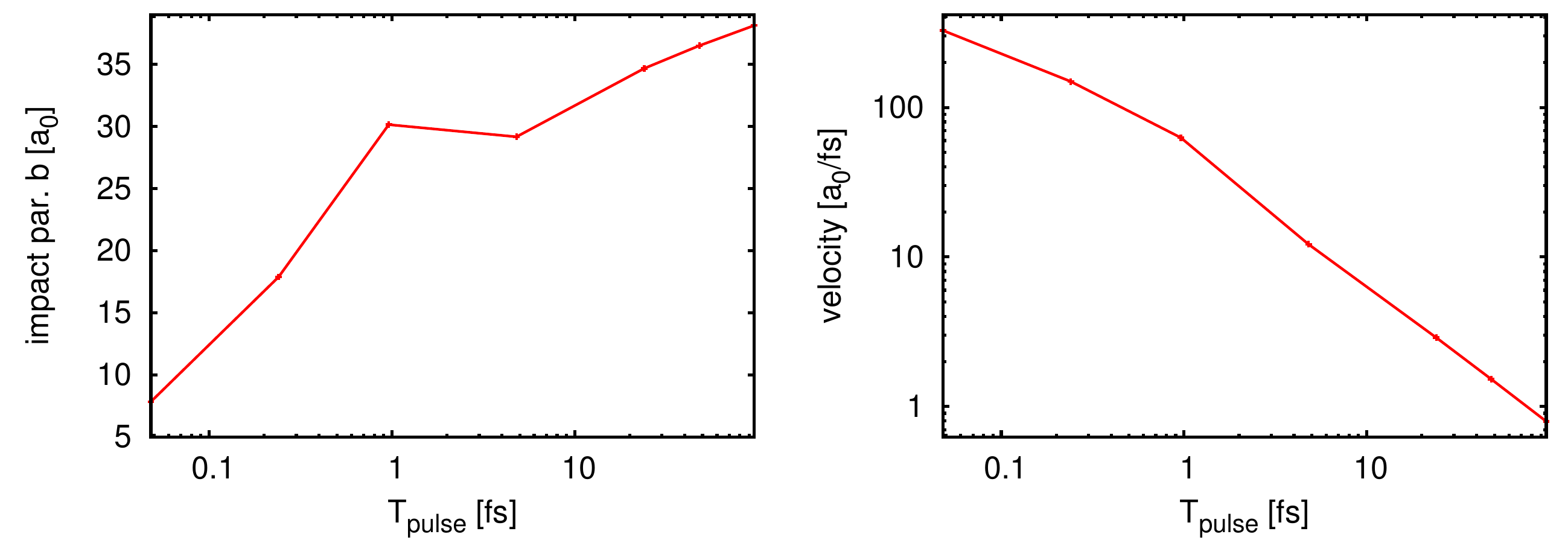}}
\vspace*{0.5em}
\centerline{\includegraphics[width=\linewidth]{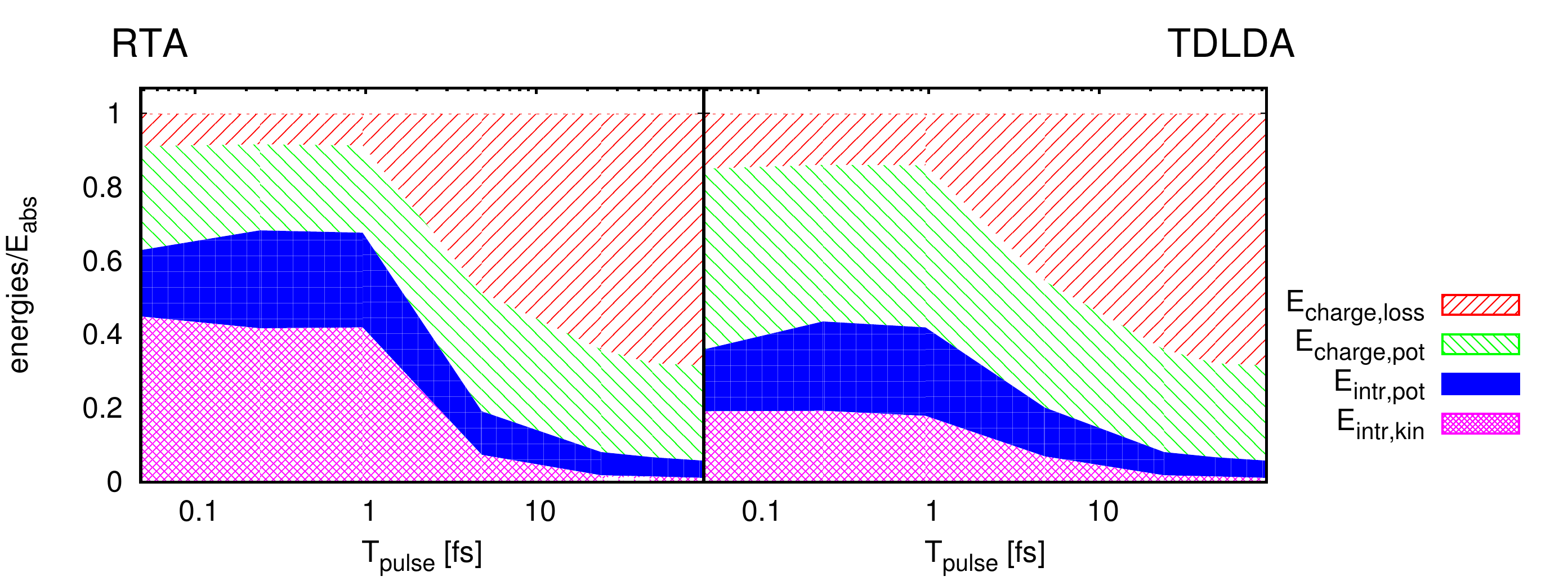}}
\caption{\label{fig:na40_pulse_balance2} Lower panel: Energy balance
  as function of pulse length $T_\mathrm{pulse}$ for Na$_{40}$ excited
  through a bypassing ion.  The impact parameter $b$ is tuned to
  provide $E_\mathrm{abs}^\mathrm{(TDLDA)}\approx 8.1$ eV.  
Upper panel: Pulse length
  $T_\mathrm{pulse}$ and excitation strength can be translated to an
  impact parameter $b$ (upper left) and ion velocity $v$ (upper
  right). This is done here for an Ar ion with charge $Q=8$. 
}
\end{figure}
An alternative excitation mechanism is collision through a by-passing
ion. We simulate that by a single dipole pulse Eq. (\ref{eq:Elaser}) with
frequency $\omega=0$. The result is shown in the lower panel of figure
\ref{fig:na40_pulse_balance2}. There are clearly
two very different regimes. For $T_\mathrm{pulse}\leq{1}$ fs, we
encounter practically an instantaneous excitation by a Dirac $\delta$
pulse, practically a boost. Here, the relations between ionization and
intrinsic energy are similar to laser excitation in the multi-photon
regime (frequencies below IP), see figure
\ref{fig:na40_E60_observables}. Much different looks the regime of
very slow ions (large $T_\mathrm{pulse}$). The intrinsic excitation
shrinks dramatically. Almost all energy flows into ionization. The
efficiency of ionization is here even better than for the one-photon
regime (high frequencies) in figure \ref{fig:na40_E60_observables}. Thus we
can conclude that collision by very slow, highly charged ions is the
softest way of ionization. 

The field exerted by a highly charged ion passing by was simulated for
simplicity by a single, zero frequency pulse. This can be translated
into collision parameters. We have done that for an Ar ion with charge
$Q=8$ as example. The peak field strength $E_0$ is related to the
impact parameters $b$ as $E_0=8Q/b^2$ and the passing time is
identified as the FWHM of field strength in the pulse which yields an
estimate for the velocity as $v=2b/T_\mathrm{pulse}$. The result of
this identification for fixed excitation energy
  $E_\mathrm{abs}=8.1$ eV is shown in the upper panel of figure
\ref{fig:na40_pulse_balance2}. The sample of $T_\mathrm{pulse}$
produces a huge span of collisional conditions.

A word is in order about the ``ideal case'' of slow collisions. It may
be not as ideal as it looks at first glance. Mind that the impact
parameter $b$ cannot be controlled in a collision. We encounter always
a mix of impact parameters thus leaving clusters in very different
excitation stages. A fair investigation has to produce the whole
excitation cross section, integrated over all impact parameters. Only
then we can judge finally whether slow collisions are a good means for
cold ionization.

\section{Conclusion}

In this paper, we have investigated from a theoretical perspective the
effect of dissipation on the energy balance in metal clusters under
the influence of strong electro-magnetic pulses. Particular attention
was paid to the branching between thermalization (intrinsic energy)
and ionization (energy export by electron emission). Basis of the
description was time-dependent density functional theory at the level
of the Time-Dependent Local-Density Approximation (TDLDA) augmented by
an averaged self-interaction correction. For a pertinent description
of dissipation, we include also dynamical correlations using the
Relaxation-Time Approximation (RTA). Test cases are Na clusters,
mainly Na$_{40}$ complemented by a few cases with different size and
charge state.

We have investigated laser excitation looking at the dependence of
energy balance on the main laser parameters, frequency, intensity
(field strength), and pulse length. Frequency is found to be the most
critical parameter. Dissipation is much more important for resonant
excitation than for non-resonant cases. It takes away energy from the
coherent dipole oscillations induced from the laser field and converts
it to intrinsic energy. This, in turn, reduces the energy loss by
induced emission and so enhances significantly the energy absorption
from the laser field. The effect continues steadily and thus grows
huge the longer the laser pulse.  Another crucial mark is set by
ionization threshold. For frequencies below, the fraction of intrinsic
excitation is generally larger than for frequencies above. Direct
emission (one-photon processes) is fast and leaves dissipation no
chance. Thus dissipative effects are negligible for high frequencies and
RTA behaves almost identical with TDLDA. The other two laser
parameters, intensity and pulse length shows much less dramatic
trends in the energy balance. Noteworthy are here two effects. First, 
the dissipative enhancement of energy absorption in the resonant case
increases linearly with pulse length. Second, with increasing
intensity (field strength), the transition from the frequency
dominated to the field dominated regime drives the energy balance to
become more similar for the different frequencies (i.e. independent of
frequency). Field emission in the strong field regime comes along with
producing less intrinsic energy.

The impact of system charge and system size was investigated for
simplicity with an instantaneous dipole boost excitation. The charge
state of a cluster changes systematically the relation between
electron emission and intrinsic heating in an obvious manner: the higher
the charge, the harder it becomes to emit an electron and thus a
larger fraction of the absorbed energy is kept in the cluster and
converted to intrinsic energy. Effects of cluster size are weak.
Shell structure still plays a role for small excitations and
becomes unimportant for higher energies.

We have also investigated excitation by a highly charged ion passing
by the cluster. There is a dramatic change of energy balance with
impact parameter. Close collisions exert a short pulse which leads to 
significant intrinsic energy (more than 50\%) if dissipation is
accounted for. Distant collisions soak off electrons very gently and
achieve high ionization while depositing very little intrinsic
energy. 

The trends of the energy balance with pulse profile and pulse
parameters are all plausible. It is interesting to check these effects
for other systems (bonding types, geometries). Research in this
direction is underway.

\section*{Acknowledgments}
This work was supported by the CNRS and the Midi-Pyr\'en\'ees region
(doctoral allocation number 13050239), and the Institut Universitaire
de France. It was granted access to the HPC resources of IDRIS under
the allocation 2014--095115 made by GENCI (Grand Equipement National
de Calcul Intensif), of CalMiP (Calcul en Midi-Pyr\'en\'ees) under the
allocation P1238, and to the Regionales Rechenzentrum Erlangen (RRZE)
of the Friedrich-Alexander university Erlangen/N\"urnberg.


\appendix

\section{Boundary correction to laser energy}
\label{app:boundcorr}

Starting point for the computation of the energy absorbed from an
external laser field is the definition in terms of the current
$\mathbf{j}$ which reads
\begin{equation}
  E_\mathrm{abs}^\mathrm{(j)}(t)
  =
  \int_0^t dt'\int d^3r\,\mathbf{E}_0(t')\cdot\mathbf{j}(\mathbf{r},t')
\end{equation}
This is turned, by virtue of the continuity equation
$\partial_t\rho=\nabla\cdot\mathbf{j}$, into an expression in terms of
$\partial_t\rho$, namely:
\begin{equation}
  E_\mathrm{abs}^\mathrm{(\rho)}(t)
  =
  \int_0^t dt'\int d^3r\,\mathbf{E}_0(t')\cdot\mathbf{r}\,\partial_t\rho(\mathbf{r},t')
\end{equation}
This form is easier to evaluate because $\rho$ is readily available
while $\mathbf{j}$ needs to be computed separately. The problem is
that the continuity equation holds only for Hermitian propagation of
the s.p. wavefunctions. To be more specific, we have to write
\begin{equation}
  \partial_t\rho_\mathrm{herm}
  =
  \nabla\cdot\mathbf{j}
\end{equation}
where $\partial_t\rho_\mathrm{herm}$ is the part stemming from
Hermitian propagation
$\partial_t\psi_\alpha=\left[\hat{h},\psi_\alpha\right]$.  Absorbing
boundaries introduce a non-Hermitian contribution to time evolution
and so spoil the continuity equation for the total density.
Subsequently, the relation
$E_\mathrm{abs}^\mathrm{(j)}=E_\mathrm{abs}^\mathrm{(\rho)}$ is not
guaranteed any more. But we can split the time-derivative of total density
 $\partial_t\rho$ into Hermitian part and contribution from
absorbing bounds as
\begin{eqnarray}
  \partial_t\rho_\mathrm{herm}
  &=&
  \partial_t\rho
  -
  \partial_t\rho_\mathrm{mask}
  \quad,
\label{eq:separ}
\\
  \partial_t\rho_\mathrm{mask}
  &=&
  \frac{1-\mathcal{M}^2}{\delta t}\,\sum_\alpha\left|\psi_\alpha\right|^2
  \quad
\end{eqnarray}
where $\mathcal{M}$ is the mask function and ${\delta t}$ the size of
the time step. This separation Eq. (\ref{eq:separ}) allows to repair the
relation $E_\mathrm{abs}^\mathrm{(j)}$ and
$E_\mathrm{abs}^\mathrm{(\rho)}$ as
\begin{eqnarray}
  E_\mathrm{abs}^\mathrm{(j)}(t)
  &=&
  E_\mathrm{abs}^\mathrm{(\rho)}
  -
  E_\mathrm{abs}^\mathrm{(mask)}
\\
  E_\mathrm{abs}^\mathrm{(mask)}
  &=&
  \int_0^t dt'\,\mathbf{E}_0(t')\cdot\mathbf{r}\,\partial_t\rho_\mathrm{mask}(\mathbf{r},t')
  \quad.
\end{eqnarray}

\section{On the semi-classical intrinsic energy}
\label{app:semicl}

\begin{figure}
\centerline{\includegraphics[width=\linewidth]{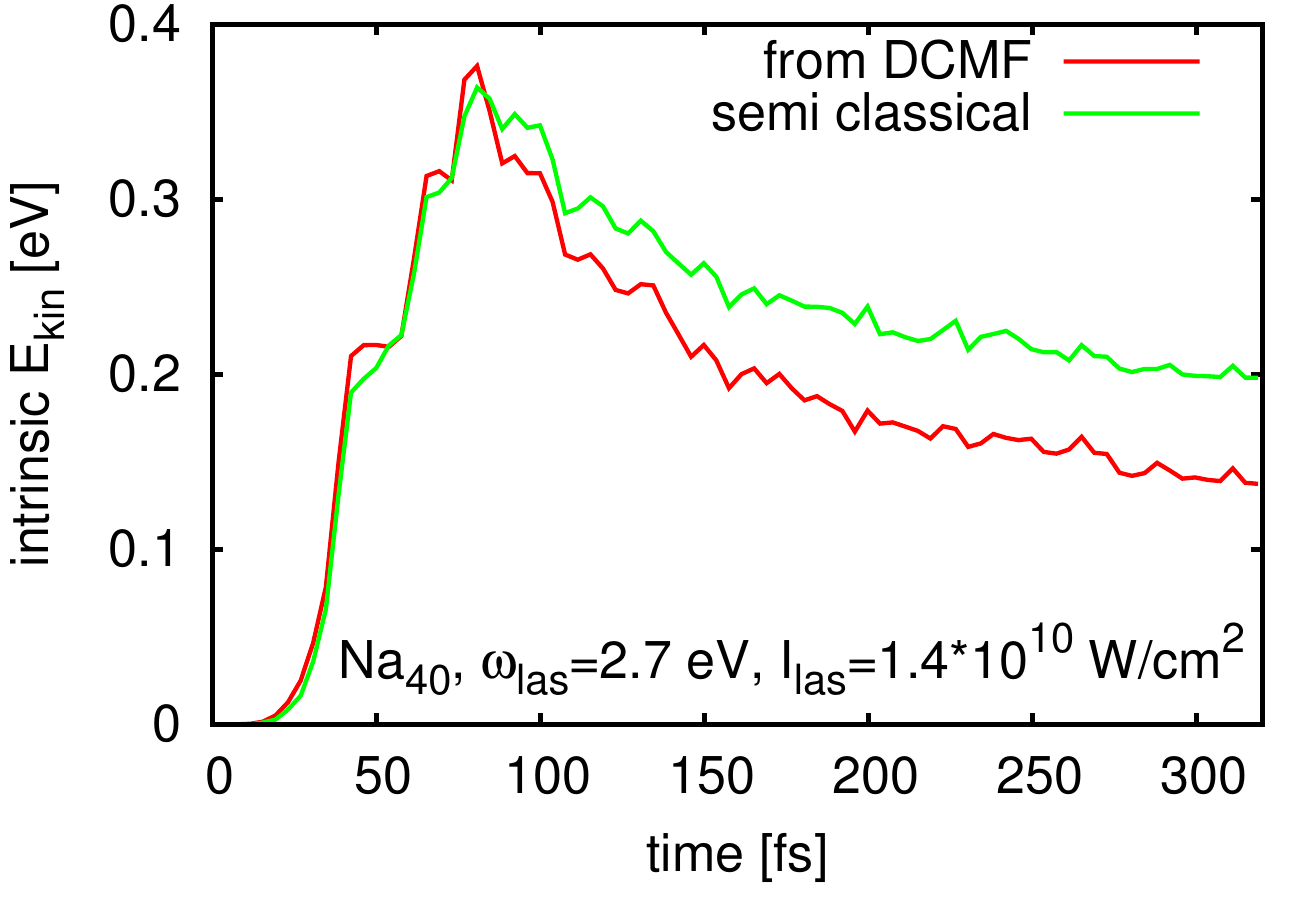}}
\caption{\label{fig:na40_o20_e0013_compEintr} Comparison of intrinsic
  kinetic energy $E_\mathrm{intr,kin}$ from fully quantum-mechanical
  DCMF definition (\ref{eq:EintrDCMF}) with the semi-classical
  estimate $E_\mathrm{intr,kin}^\mathrm{(ETF)}$ for the example of
  Na$_{40}$ excited by a laser pulse with $\omega_\mathrm{las}=2.7$
  eV, $I_\mathrm{las}=1.4\times 10^{10}$ W/cm$^2$, and
  $T_\mathrm{pulse}=96$ fs.
}
\end{figure}
The fully quantum-mechanical definition Eq. (\ref{eq:EintrDCMF}) of an
intrinsic kinetic energy employs a DCMF iteration which is naturally
available when propagating with RTA but becomes a rather expensive
extra step in pure TDLDA. Thus one often sidesteps to a simpler
semi-classical estimate from the extended Thomas-Fermi approach
\cite{Bra97a}
\begin{eqnarray*}
  E_\mathrm{intr,kin}^\mathrm{(ETF)}
  &=&
  E_\mathrm{kin}^\mathrm{(TDLDA)}
\nonumber \\
  &&
  -
  \int d^3r\left(
   {\textstyle\frac{2}{3}}(3\pi^2)^{2/3}\rho^{2/3}
   +
   \frac{(\nabla\rho)^2}{18\rho}
  \right)
  -
  E_\mathrm{coll}
\end{eqnarray*}
with the collective energy from Eq. (\ref{eq:Ecoll}).  The two
definitions are compared in figure
\ref{fig:na40_o20_e0013_compEintr}. The semi-classical
$E_\mathrm{intr,kin}^\mathrm{(ETF)}$ is a robust order-of-magnitude
estimate which works particularly well in the early phases of
excitation. 

The case is more involved than it appears in figure
\ref{fig:na40_o20_e0013_compEintr}. Actually, the mismatch starts at
$t=0$. But we shift the value of $E_\mathrm{intr,kin}^\mathrm{(ETF)}$
to match at $t=0$, precisely because it is a semi classical estimate,
thus not fully vanishing in ground state. The punishment is then a
mismatch at large times. This may have to be discussed.

\bibliographystyle{epj}
\bibliography{relax-syst}

\begin{thebibliography}{66}

\bibitem{Gro90}
E.K.U. Gross, W.~Kohn, Adv. Quant. Chem. \textbf{21}, 255 (1990)

\bibitem{Gro96}
E.K.U. Gross, J.F. Dobson, M.~Petersilka, Top. Curr. Chem. \textbf{181}, 81
  (1996)

\bibitem{Mar12}
M.A.L. Marques, N.T. Maitra, F.M.S. Nogueira, E.K.U. Gross, A.~Rubio,
  \emph{{Fundamentals of Time-Dependent Density Functional Theory }} (Lect.
  Notes in Phys. vol 837,Springer-Verlag, Berlin, 2012)

\bibitem{Neg82aR}
J.W. Negele, Rev. Mod. Phys. \textbf{54}, 913 (1982)

\bibitem{Dav85a}
K.T.R. Davies, K.R.S. Devi, S.E. Koonin, M.R. Strayer, in \emph{Treatise on
  Heavy--Ion Physics, Vol. 3 Compound System Phenomena}, edited by D.A. Bromley
  (Plenum Press, New York, 1985), p.~3

\bibitem{Ben03aR}
M.~Bender, P.H. Heenen, P.G. Reinhard, Rev. Mod. Phys. \textbf{75}, 121 (2003)

\bibitem{Ber94aB}
G.F. Bertsch, R.~Broglia, \emph{Oscillations in Finite Quantum Systems}
  (Cambridge University Press, Cambridge, 1994)

\bibitem{Rei06aR}
P.G. Reinhard, E.~Suraud, in \emph{Time-dependent density functional theory},
  edited by M.A.L. Marques, C.A. Ullrich, F.~Nogueira (Springer, Berlin, 2006),
  Vol. 706 of \emph{Lecture Notes in Physics}, p. 391

\bibitem{Fen10}
T.~Fennel, , K.H. Meiwes-Broer, J.~Tiggesb\"aumker, P.G. Reinhard, P.M. Dinh,
  E.~Suraud, Rev. Mod. Phys. \textbf{82}, 1793 (2010)

\bibitem{golabek2009}
{C\'edric Golabek}, {C\'edric Simenel}, Phys. Rev. Lett. \textbf{103}(4),
  042701 (2009), ISSN 0031-9007

\bibitem{Obe10a}
V.E. Oberacker, A.S. Umar, J.A. Maruhn, P.G. Reinhard, Phys. Rev. C
  \textbf{82}, 034603 (2010), http://arxiv.org/abs/1007.4774,
  \texttt{http://link.aps.org/doi/10.1103/PhysRevC.82.034603}

\bibitem{Rei94aR}
P.G. Reinhard, C.~Toepffer, Int. J. Mod. Phys. E \textbf{3}, 435 (1994)

\bibitem{Ber83aR}
G.F. Bertsch, P.F. Bortignon, R.A. Broglia, Rev. Mod. Phys. \textbf{55}, 287
  (1983)

\bibitem{Abe96}
Y.~Abe, S.~Ayik, P.G. Reinhard, E.~Suraud, Phys. Rep. \textbf{275}, 49 (1996)

\bibitem{Sar12a}
V.V. Sargsyan, G.G. Adamian, N.V. Antonenko, W.~Scheid, H.Q. Zhang, Phys. Rev.
  C \textbf{85}, 024616 (2012), [Erratum: Phys. Rev.C85,069903(2012)]

\bibitem{Lac14a}
D.~Lacroix, S.~Ayik, Eur. Phys. J. A \textbf{50}(6), 95 (2014)

\bibitem{Del98a}
N.D. Fatti, R.~Bouffanais, F.~Vallée, C.~Flytzanis, Phys. Rev. Lett.
  \textbf{81}, 922 (1998)

\bibitem{Voi00}
C.~Voisin, D.~Christofilos, N.D. Fatti, F.~Vall\'ee, B.~Pr\"ovel, E.~Cottancin,
  J.~Lerm\'e, M.~Pellarin, M.~Broyer, Phys. Rev. Lett. \textbf{85}, 2200 (2000)

\bibitem{Nae97}
U.~N\"aher, S.~Bj\"ornholm, S.~Frauendorf, F.~Garcias, C.~Guet, Phys. Rep.
  \textbf{285}, 245 (1997)

\bibitem{Cam00}
E.E.B. Campbell, K.~Hansen, K.~Hoffmann, G.~Korn, M.~Tchaplyguine, M.~Wittmann,
  I.V. Hertel, Phys. Rev. Lett. \textbf{84}, 2128 (2000)

\bibitem{Sch01c}
M.~Schmidt, R.~Kusche, T.~Hippler, J.~Donges, W.~Kronm\"uller, B.~v~Issendorff,
  H.~Haberland, Phys. Rev. Lett. \textbf{86}, 1191 (2001)

\bibitem{Feh06b}
F.~Fehrer, P.G. Reinhard, E.~Suraud, Appl. Phys. A \textbf{82}, 145 (2006)

\bibitem{Kje10}
M.~Kjellberg, O.~Johansson, F.~Jonsson, A.V. Bulgakov, C.~Bordas, E.E.B.
  Campbell, K.~Hansen, Phys. Rev. A \textbf{81}, 023202 (2010)

\bibitem{Rei15d}
P.G. Reinhard, E.~Suraud, Ann. Phys. (N.Y.) \textbf{354}, 183 (2015),
  \texttt{http://dx.doi.org/10.1016/j.aop.2014.12.01}

\bibitem{Rei98aB}
L.E. Reichl, \emph{A Modern Course in Statistical Physics} (Wiley, New York,
  1998)

\bibitem{Cer88}
C.~Cercignani, \emph{{The Boltzmann equation and its applications}} (Applied
  Mathematical Sciences 67, Springer, New York, 1988)

\bibitem{Ueh33}
E.A. Uehling, G.E. Uhlenbeck, Phys. Rev. \textbf{43}, 552 (1933)

\bibitem{Ber88}
G.F. Bertsch, S.~{Das Gupta}, Phys. Rep. \textbf{160}, 190 (1988)

\bibitem{Dur00}
D.~Durand, E.~Suraud, B.~Tamain, \emph{{Nuclear Dynamics in the Nucleonic
  Regime}} (Institute of Physics, London, 2000)

\bibitem{Dom98b}
A.~Domps, P.G. Reinhard, E.~Suraud, Phys. Rev. Lett. \textbf{81}, 5524 (1998)

\bibitem{Fen04}
T.~Fennel, G.F. Bertsch, K.H. Meiwes-Broer, Eur. Phys. J. D \textbf{29}, 367
  (2004)

\bibitem{Kad62}
L.P. Kadanoff, G.~Baym, \emph{{Quantum Statistical Mechanics: Green's Function
  Methods in Equilibrium and Nonequilibrium Problems}} (Frontiers in physics,
  Benjamin, New York, 1962)

\bibitem{Bha54}
P.L. Bhatnagar, E.P. Gross, M.~Krock, Phys. Rev. \textbf{94}, 1954 (511)

\bibitem{Ash76}
N.W. Ashcroft, N.D. Mermin, \emph{{Solid State Physics}} (Saunders College,
  Philadelphia, 1976)

\bibitem{Pin66}
D.~Pines, P.~Nozi\`eres, \emph{{The Theory of Quantum Liquids}} (W A Benjamin,
  New York, 1966)

\bibitem{Dut12a}
A.~Dutta, C.~Trefzger, K.~Sengupta, Phys. Rev. B \textbf{86}, 085140 (2012)

\bibitem{Kra07a}
P.~Krause, T.~Klamroth, P.~Saalfrank, J. Chem. Phys. \textbf{127}, 034107
  (2007)

\bibitem{Rei92b}
P.G. Reinhard, E.~Suraud, Ann. Phys. (N.Y.) \textbf{216}, 98 (1992)

\bibitem{Sur14d}
E.~Suraud, P.G. Reinhard, New J. Phys. \textbf{16}, 063066 (2014),
  \texttt{http://stacks.iop.org/1367-2630/16/063066}

\bibitem{Lac16a}
L.~Lacombe, P.G. Reinhard, P.M. Dinh, E.~Suraud, J. Phys. B \textbf{49}, 245101
  (2016), \texttt{doi:10.1088/0953-4075/49/24/245101}

\bibitem{Leg02}
C.~Legrand, E.~Suraud, P.G. Reinhard, J. Phys. B \textbf{35}, 1115 (2002)

\bibitem{Klu13}
P.~Kl\"upfel, P.M. Dinh, P.G. Reinhard, E.~Suraud, Phys. Rev. A \textbf{88},
  052501 (2013)

\bibitem{Cal00}
F.~Calvayrac, P.G. Reinhard, E.~Suraud, C.A. Ullrich, Phys. Rep. \textbf{337},
  493 (2000)

\bibitem{Rei04aB}
P.G. Reinhard, E.~Suraud, \emph{Introduction to Cluster Dynamics} (Wiley, New
  York, 2004)

\bibitem{Kue99}
S.~K\"ummel, M.~Brack, P.G. Reinhard, Eur. Phys. J. D \textbf{9}, 149 (1999)

\bibitem{Per92}
J.P. Perdew, Y.~Wang, Phys. Rev. B \textbf{45}, 13244 (1992)

\bibitem{Mon94a}
B.~Montag, P.G. Reinhard, Phys. Lett. A \textbf{193}, 380 (1994)

\bibitem{Mon95a}
B.~Montag, P.G. Reinhard, Z. Phys. D \textbf{33}, 265 (1995)

\bibitem{Dav81a}
K.T.R. Davies, S.E. Koonin, Phys. Rev. \textbf{C23}, 2042 (1981)

\bibitem{Fei82}
M.D. Feit, J.A. Fleck, A.~Steiger, J. Comp. Phys. \textbf{47}, 412 (1982)

\bibitem{Blu92}
V.~Blum, G.~Lauritsch, J.A. Maruhn, P.G. Reinhard, J. Comp. Phys \textbf{100},
  364 (1992)

\bibitem{Rei06c}
P.G. Reinhard, P.D. Stevenson, D.~Almehed, J.A. Maruhn, M.R. Strayer, Phys.
  Rev. E \textbf{73}, 036709 (2006)

\bibitem{Dre90}
R.M. Dreizler, E.K.U. Gross, \emph{{Density Functional Theory: An Approach to
  the Quantum Many-Body Problem}} (Springer-Verlag, Berlin, 1990)

\bibitem{Koe08a}
J.~K\"ohn, R.~Redmer, K.H. Meiwes-Broer, T.~Fennel, Phys. Rev. A \textbf{77},
  033202 (2008)

\bibitem{Koe12a}
J.~K\"ohn, R.~Redmer, T.~Fennel, New J. Phys. \textbf{14}, 055011 (2012)

\bibitem{Cus85a}
R.~Cusson, P.G. Reinhard, J.~Maruhn, W.~Greiner, M.~Strayer, Z. Phys. A
  \textbf{320}, 475 (1985)

\bibitem{Cha95}
F.~Chandezon, C.~Guet, B.A. Huber, D.~Jalabert, M.~Maurel, E.~Monnand,
  C.~Ristori, J.C. Rocco, Phys. Rev. Lett. \textbf{74}, 3784 (1995)

\bibitem{Rei96b}
P.G. Reinhard, O.~Genzken, M.~Brack, Ann. Phys. (Leipzig) \textbf{5}, 1 (1996)

\bibitem{Bab97}
J.~Babst, P.G. Reinhard, Z. Phys. D \textbf{42}, 209 (1997)

\bibitem{Lou09aB}
R.~Loudon, \emph{The Quantum Theory of Light} (Oxford Science Publications,
  Oxford, 2009)

\bibitem{Kel65}
L.V. Keldysh, Sov. Phys. JETP \textbf{20}, 1307 (1965)

\bibitem{Rei99a}
P.G. Reinhard, F.~Calvayrac, C.~Kohl, S.~K\"ummel, E.~Suraud, C.A. Ullrich,
  M.~Brack, Eur. Phys. J. D \textbf{9}, 111 (1999)

\bibitem{Mon94b}
B.~Montag, P.G. Reinhard, J.~Meyer, Z. Phys. D \textbf{32}, 125 (1994)

\bibitem{Bra93}
M.~Brack, Rev. Mod. Phys. \textbf{65}, 677 (1993)

\bibitem{Mar93}
T.P. Martin, Phys. Rep. \textbf{273}, 199 (1993)

\bibitem{Bra97a}
M.~Brack, R.K. Bhaduri, \emph{{Semiclassical Physics}} (Addision-Wesley,
  Reading, 1997)

\end{thebibliography}

\end{document}